\documentstyle[12pt]{article}
\begin{document}

\begin{center}
{\Large {\bf Antiferromagnetism and Superconductivity in a Model
with Extended Pairing Interactions}}\\
\vspace{1.0cm} 
{{\bf Tulika Maitra$^{*}$,}\footnote{email: tulika@phy.iitkgp.ernet.in}  
{\bf H. Beck}$^{**}$\footnote{email: Hans.Beck@unine.ch}
and {\bf A. Taraphder}$^{*\dagger}$\footnote{email: arghya@phy.iitkgp.ernet.in, 
arghya@cts.iitkgp.ernet.in}}\\ 

\noindent $^{*}$Department of Physics \& Meteorology and  
$^{\dagger}$Centre for Theoretical Studies, \\
Indian Institute of Technology, Kharagpur 721302 India \\   
\noindent $^{**}$ University of Neuchatel - rue de Breguet 1, 2000 Neuchatel,
Switzerland.  
\end{center}  
\begin{abstract}
\vspace{.5cm} 

The competition between antiferromagnetism and the $d+id$ superconducting state
is studied in a model with near and next near neighbour interactions in the
absence of any on-site repulsion. A mean field study shows that it is
possible to have simultaneous occurrence of an antiferromagnetic and a singlet 
$d+id$ superconducting state in this model. In addition, such a coexistence 
generates a triplet $d+id$ superconducting order parameter with centre of 
mass momentum $Q=(\pi, \pi)$ {\it dynamically} having the same orbital 
symmetry as the singlet superconductor. Inclusion of next nearest neighbour 
hopping in the band stabilises the $d_{xy}$ superconducting state away from 
half filling, the topology of the phase diagram, though, remains similar to 
the near neighbour model. 
In view of the very recent observation of a broad region of coexistence of 
antiferrmagnetic and unconventional superconducting states in organic 
superconductors, the possibility of observation of the triplet state 
has been outlined.

\end{abstract}  
\vspace{.5cm} 

\noindent {\bf I. Introduction}  
\vspace{.5cm} 

Interests on the interplay between antiferromagnetism and superconductivity 
date back quite a while as in certain organic and heavy fermion 
superconductors, superconductivity is known to coexist with an 
antiferromagnetic (AF) phase at low temperatures\cite{bour,hews}. In a recent 
study of the organic superconductor $\kappa$-(ET)$_{2}$Cu[N(CN)$_{2}$]Cl 
the most complete phase diagram has been obtained as a function of pressure 
and a region of coexistence of unconventional SC and AF LRO is 
observed\cite{lefeb}. 
In the high T$_c$ cuprates, the superconducting state abuts (albeit 
with a small gap) the AF state and recent inelastic neutron 
scattering\cite{rosaM,bourges} reveals considerable AF fluctuations 
deep inside the superconducting (SC) state in YBCO. 
There are, actually, quite a few similarities between the 
organic and high T$_c$ superconductors\cite{mcken}. Indeed, 
antiferromagnetism, in some theories, is considered to lie at the heart 
of the mechanism that drives unconventional superconductivity\cite{pines}. 

Several investigations were carried out\cite{machida} to study the
phase diagram and nature of phase transitions between the AF state
and the SC state in the context of the organic and heavy fermion
superconductors. Meintrup et al.\cite{mein} proposed a model 
in the general context of antiferromagnetism and superconductivity
in which nearest neighbour singlet pairing interaction was shown
to accommodate both SC and AF states and their coexistence in certain
range of parameters. These authors, and several others before 
them\cite{sch,micnas},
showed that it is not necessary to have strong on-site repulsion
to generate AF LRO. Correlated models with extended range of interactions
can produce AF LRO and SC as well. 

In a recently proposed SO(5) theory of superconductivity Zhang\cite{zhang}
considered a five dimensional order parameter space $(\psi_{1},\cdots, 
\psi_{5})$ with $\psi_{1}+i\psi_{5}$ being the SC order parameter and
the remaining three constituting the AF moment. A rotation in this five 
dimensional order parameter space, effected by the spin 1, charge 2, 
$\pi-$operators, ($\pi^{\dagger}_{\alpha}=\sum_{k}g(k)c^{\dagger}_{k+Q\uparrow}
c^{\dagger}_{-k\downarrow}$, where $c^{\dagger}(c)$'s are the electron 
creation (destruction) operators and $Q=(\pi,\pi)$ in d=2, leads to 
the transition from the AF to the SC state. In this theory, the triplet 
magnetic excitation of the quantum disordered phase is identified with this 
$\pi$-triplet mode in the SC phase. In an exact diagonalization 
study of the $t-J$ model\cite{meixner}, the dynamical correlation functions of 
the $\pi-$operators have been calculated and found to be non-zero. The 
existence of both AF and triplet pairing amplitude with net momentum $Q$ 
was reported earlier in the mean-field study of a pairing Hamiltonian
in the context of the heavy fermions and organic superconductors
\cite{machida,fent}, although the conditions under which the triplet amplitude
appears and the modifications of the phase boundaries due to this triplet 
amplitude were not dealt with. In a recent investigation, Kyung\cite{kyung} 
considered explicit mean-field pairing interactions in the singlet and 
triplet channel with a repulsive on-site interaction to stabilize the $d$-wave 
SC state (over $s$-wave) and discussed the coexistence of a dynamically 
generated triplet SC pair amplitude and AF long range order (LRO). In 
this case, superconductivity is governed by the attractive interactions 
in the appropriate channels while the AF state owes its origin primarily 
to the on-site repulsion in the usual manner.  

We start from a Hamiltonian with nearest and next nearest neighbour interaction
and consider pairing in the $d$-wave and $d+id$-wave in the singlet
(and later on in the triplet) channel along with the antiferromagnetic LRO. 
The on-site interaction is assumed to be small\cite{scal} (set to zero here) 
and both the LRO and off diagonal long range order (ODLRO) are governed by 
the same interactions in a manner similar to the case studied by Meintrup 
et. al.\cite{mein} (where the on-site interaction was absent as well). 
Using a mean-field analysis, these authors studied the coexistence of different
singlet SC and AF LRO states in their model. Related extended range models
have been studied by numerical methods\cite{sch}  
and mean-field\cite{micnas} theory earlier, although a detailed study with the
possibility of several SC symmetries and AF order have not been undertaken.

There have been suggestions
\cite{tesa} for the existence of $d+id$ state in the high $T_c$ systems
followed by possible observation in a series of experiments\cite{kris,aubin}.
Recently, it has been shown\cite{sach} from a renormalization group analysis
of the fluctuations that the transition to $d+id$ state possesses a stable 
fixed point. Kino and Fukuyama\cite{kino} considered a model with only on-site
repulsion for the organic superconductors in the intermediate coupling
range (typically the on-site interaction is about half the band width). Such
a model, although accounts for the AF phase and the metal-insulator transition,
fails to explain the large SC phase observed in\cite{lefeb}. Extended range 
attractive interactions with the right symmetry are necessary to obtain these
unconventional SC states. Additional processes discussed in ref.\cite{scal}, 
particularly in the large metallic region above the SC phase,
possibly reduce the on-site interaction in the organic systems further.

A preliminary report of the coexistence of and competition between the 
singlet superconducting
$d_{x^2-y^2}$, $d_{xy}$,  $d_{x^2-y^2}+id_{xy}$ (the so called $d+id$ state) 
states and AF LRO in a model similar to that of Meintrup et. al. with extended
range of interaction has been presented recently by two of us\cite{beck}. 
We extend this calculation and show in the present work that in the presence of
such coexisting singlet SC order parameter and AF LRO, a triplet pairing 
amplitude with centre of mass momentum $Q$ is dynamically generated
even if there is no explicit interaction in the Hamiltonian in that channel.
It is not a-priori obvious that the dynamical generation of the triplet 
amplitude should occur in a model where the AF and SC states are governed by 
combinations of the same interactions. In the model considered by 
Kyung\cite{kyung} the relative strength of these two competing states are 
governed primarily by separate and independent interaction parameters. 
We also figure out how the phase diagram gets modified in the presence of 
the triplet pairing amplitude in the present model. In section II we discuss
the model under consideration. Section III concerns with the results,  
discussion and concluding remarks. 
\vspace{.4cm} 

\noindent {\bf II. Model and Calculations} 
\vspace{.4cm} 

\noindent The model studied here incorporates antiferrmagnetic LRO 
and superconductivity  and is given by the Hamiltonian\cite{beck}
$${\bf H}=\sum_{k,\sigma} \xi_{\bf k}c_{{\bf k}\sigma}^{\dag}
c_{{\bf k}\sigma}+\sum_{<ij>\sigma}V_{ij}n_{i\sigma}n_{j-\sigma} + 
\sum_{<<ij>>\sigma\sigma^{\prime}}{\tilde V}_{ij}n_{i\sigma}n_{j\sigma^\prime}
\eqno(1)$$ 
where the sum over $<ij>$ extends over near neighbour and $<<ij>>$ over next 
near neighbour sites. We take $V_1$ and $V_2$ as the corresponding interaction 
strengths (both $V_1$ and $V_2$ are negative) and write $\xi_{\bf k}=
\epsilon_{\bf k}-\mu$. In the absence of an on-site repulsion, this
model is perhaps the simplest that produces AFM as well as superconductivity
in the $d+id$ channel.

Elementary physical reasoning shows how an AF state appears in this 
Hamiltonian. In the classical limit ($t_{ij}=0$) the Hamiltonian has 
near-neighbour attractive density-density interaction (amongst opposite spins) 
that leads to AF spin correlation (of Ising symmetry) among nearest neighbour
spins. An on-site repulsive term would have stabilised this further and 
the region of AF LRO would extend in the phase space. The second neighbour
attractive density-density correlation term is spin independent and stabilises
a $d_{xy}$ order in the quantum limit. The regions of stability of 
$d_{x^2-y^2}$, $d_{xy}$ and the $d+id$ state for the range of values 
of $V_1$ and $V_2$ have been discussed in\cite{maitra}. Extensive literature
exists for models in the opposite limit of repulsive extended range interactions
where the classical limit gives rise to charge density waves\cite{tarap}. In
the absence of $V_{2}$, Monte-Carlo calculations\cite{sch} of model (1) shows 
AF phase at all densities. At half-filling the ground state is N\'{e}el
ordered while away from half-filling there is evidence for phase separation
between AF ordered and empty domains. Mean-field analysis\cite{mein} captures
much of these features qualitatively, although a realistic description
of the phase separation eludes such calculations as expected.  

In order to use mean-field description for the symmetry-broken states we define 
the operators corresponding to the singlet and triplet SC order parameters 
in real space\cite{scal}

$$\Lambda_{i,s}=\frac{1}{4}\sum_{\delta,\sigma}\sigma c_{i+\delta,\sigma}
c_{i,-\sigma}\phi(\delta) \,\,\,\, {\rm and} \,\,\,\,  
\Lambda_{i,t}=\frac{1}{4}\sum_{\delta,\sigma} c_{i+\delta,\sigma}
c_{i,-\sigma}\phi(\delta)\eqno(2)$$ 
\noindent where $\delta$ is the usual near-neighbour translation vector and 
a choice of the form factor $\phi(\delta)$ with 
$\phi_{1}(\delta)=1$ for $\delta=(\pm 1,0)$ and  $\phi_{1}(\delta)=-1$ 
for $\delta=(0,\pm 1)$ (with $\phi_{1}(\delta)=0$ for all other choice of 
$\delta$) ensures that the SC OP has $d_{x^2-y^2}$ symmetry. For $d_{xy}$
symmetry, one takes the form factor (the only non-zero terms) $\phi$ as
$\phi_{2}(\delta)=1$ for $\delta=(\pm 1,\pm 1)$ and $\phi_{2}(\delta)=-1$ 
for $\delta=(\pm 1,\mp 1)$. The operator corresponding to the AF order 
parameter is the well known form $\sum_{\sigma}\sigma c^{\dagger}_{i,\sigma}
c_{i,\sigma}.$
Writing the AF, singlet and triplet SC order parameters as ($\sigma=\pm 1$) 

$$\sum_{\sigma}<\sigma c^{\dagger}_{i,\sigma} c_{i,\sigma}>=b_{0}
e^{i{\bf Q.r}_{i}}\eqno(3a)$$

$$\frac{1}{4}\sum_{\delta,\sigma}<\sigma c_{i+\delta,\sigma}
c_{i,-\sigma}>(\phi_{1}(\delta)+i\phi_{2}(\delta))=\Delta_{s}\eqno(3b)$$ 
\noindent and

$$\frac{1}{4}\sum_{\delta,\sigma}<c_{i+\delta,\sigma}
c_{i,-\sigma}>(\phi_{1}(\delta)+i\phi_{2}(\delta))=\Delta_{t}
e^{i{\bf Q.r}_{i}}\eqno(3c).$$ 

The SC order parameters are chosen to be of $d_{x^2-y^2}$ and $d_{xy}$ 
symmetries as the interactions $V_1$ and $V_2$ are known to 
favour\cite{sach,maitra,chatt} superconductivity in such orbital symmetries.
The presence of AF order in the model (1) has already been indicated.
The superconducting order parameter $\Delta_s$, which is a spin singlet 
with d-wave orbital symmetry, can have a non-zero value when the underlying
pairing state is spatially homogeneous. On the other hand, the order 
parameter $\Delta_t$ that we write down here should be thought of as a
non-zero expectation value of the  $\pi-$operator of the SO(5) theory and
is generated only dynamically \cite{kyung1}. It can be non-zero provided
that the expectation value of the annihilation operators in (3c) changes
sign on alternating bonds. The resulting $\Delta_t$ is thus a staggered order
parameter: it changes sign from site to site in the same way as the
antiferromagnetic order parameter (3a) does. 

Using the Hartree-Fock approximation with these order parameters and going 
over to Fourier space we get a Hamiltonian in quadratic form as 
$${\bf H}_{MF}=\sum_{k,\sigma}
(\epsilon_{\bf k}-\tilde{\mu})c_{{\bf k}\sigma}^{\dag}c_{{\bf k}\sigma}+b_m
\sum_k[(c_{{\bf k}\uparrow}^{\dag}c_{{\bf k+Q}\uparrow}-c_{{\bf k}\downarrow}
^{\dag}c_{{\bf k+Q}\downarrow})+\it {h.c.}]+\sum_k(\Delta_s^{*}({\bf k})
c_{{\bf -k}\downarrow}c_{{\bf k}\uparrow}+{\it h.c.})$$
$$+\sum_k[\Delta_t^{*}({\bf k})(c_{{\bf -k}\downarrow}c_{{\bf k+Q}\uparrow}
+c_{{\bf -k-Q}\downarrow}c_{{\bf k}\uparrow})+{\it h.c.}]\eqno(4)$$

\noindent Here $\Delta_s({\bf k})= \frac{1}{2}\Delta_1(cosk_x-cosk_y)+i\Delta_2
sink_xsink_y$ and $\Delta_t({\bf k})=\frac{1}{2}\Delta_{t1}(cosk_x-cosk_y)
+i\Delta_{t2} sink_xsink_y$ are the pairing amplitudes in the singlet and 
triplet channel respectively; $b_m=Ab_0$  with $A=zV_1+zV_2$ ($z$ is the 
coordination number).
The tight binding energy dispersion on a square lattice that we use involves
upto next-near neighbour hopping in conformity with the range of
interactions considered. The dispersion is $\epsilon_{\bf k}=-2t(cosk_x+cosk_y)
+4t^{\prime} cosk_x cosk_y$ where {\it t} is the near neighbour and 
$t^\prime$ is the next near neighbour hopping integral. Here $\tilde \mu
=\mu-nz(V_1+V_2)$ and we use $\mu$ to denote $\tilde \mu$ in the foregoing 
analysis. 

The potential and the SC order parameters are expanded\cite{chatt,legg} in the 
usual 
basis functions ($B_1$ representation of $C_{4v}$) $\eta_{1}({\bf k})=
\frac{1}{2}(cosk_x-cosk_y)$ and $\eta_{2}({\bf k})=sink_xsink_y$.  
Expanding in these bases $V({\bf k, k^\prime })=\sum_{i}
V_{i}\eta_{i}({\bf k}) \eta_{i}({\bf k^\prime })$ and $\Delta_{\bf k}= 
\sum_{\bf k^\prime} V({\bf k, k^\prime})\Gamma_{\bf k^\prime}\equiv 
\sum_{i}V_{i}\Delta_{i}\eta_{i}({\bf k})$ where $\Delta_{i}=\sum_{\bf k}
\eta_{i}(\bf k) \Gamma_{\bf k}$ (and $\Gamma_{\bf k}
=<\sum_{\sigma}c_{{\bf k}\sigma} c_{{-\bf k}\sigma}>$). $V({\bf k,k'})$ comes
from the Fourier transform of the second and third terms of the Hamiltonian 
(1) (described in detail in ref. \cite{chatt}). 

The order parameters can be treated as variational parameters with the
trial Hamiltonian $H_{MF}$. The corresponding free energy functional
is given by
$$\tilde{F}=F_{0}+<H-H_{MF}>_0\eqno(5)$$ 
\noindent where $<...>_0$ denotes average with respect to $\rho_{0}=
exp [-H_{MF}/kT]/Z_0$ and $F_{0}=-kT ln(Z_0)$.
The self-consistency equations for the order parameters are obtained by 
minimising the free-energy functional $\tilde{F}$. 

$$1=\frac{1}{2N} \sum_{\bf k}
\sum_{\gamma=+,-} A\{1+\frac{{\mu^{2}}\gamma}{\alpha(\bf k)}\}
\frac{1}{E_{\gamma}(\bf k)} tanh \frac{E_{\gamma}(\bf k)}{2T}\eqno(6)$$

$$ 1 =  \frac{1}{2N} \sum_{\bf k}\sum_{\gamma=+,-} 
\frac{V_1\eta_1^2({\bf k})}
{E_{\gamma}(\bf k)}tanh\frac{E_{\gamma}(\bf k)}{2T}\eqno(7) $$

$$ 1 =  \frac{1}{2N} \sum_{\bf k}\sum_{\gamma=+,-} 
\frac{V_2\eta_2^2({\bf k})}
{E_{\gamma}(\bf k)}tanh\frac{E_{\gamma}(\bf k)}{2T}\eqno(8) $$

$$\Delta_{t1}  = -\frac{1}{2N} \sum_{\bf k}\sum_{\gamma=+,-}
{\frac{\gamma}{\alpha(\bf k)}
V_1\Delta_1\eta_1^2({\bf k})\mu Ab_0 }
\frac{1}{E_{\gamma}(\bf k)}tanh\frac{E_{\gamma}(\bf k)}{2T}\eqno(9)$$

$$\Delta_{t2}  =  -\frac{1}{2N} \sum_{\bf k}\sum_{\gamma=+,-}
{\frac{\gamma}{\alpha(\bf k)}
V_2\Delta_2\eta_2^2({\bf k})\mu Ab_0 }
\frac{1}{E_{\gamma}(\bf k)}tanh\frac{E_{\gamma}(\bf k)}{2T}\eqno(10)$$

The self-consistency equations (6)-(10) provide a set of five coupled
equations to be solved numerically. What is interesting is that although
we have not included any pairing interaction in the triplet channel,
the triplet amplitudes $\Delta_{t1}$ and $\Delta_{t2}$ are non-zero. The
simultaneous co-existence of $\Delta_{1,2}$ and $b_{0}$ ensures the existence
of this non zero amplitude.

One could, of course, include additional pairing interactions $W_1$ and $W_2$
explicitly in the Hamiltonian (1) in the triplet channels with preferred
symmetries and obtain the self-consistency equations. (An alternative and
quite commonly used approach is to include such terms in the mean-field
Hamiltonian (4) as in Kato and Machida\cite{machida} and Kyung\cite{kyung},
rather than derive them from microscopic interactions). 
This would modify all the equations (6)-(10) but the nature of the phase
diagrams remains qualitatively similar (discussed later). For the sake of
completeness, we write down the equations for the triplet amplitudes
including $W_1$ and $W_2$ and note that they reduce to (9) and (10)
without $W_{1,2}$. 

$$\Delta_{t1}  =  \frac{1}{2N} \sum_{\bf k}\sum_{\gamma=+,-}
\left \{  W_1\Delta_{t1}\eta_1^2({\bf k})+\frac{\gamma}{\alpha(\bf k)}
V_1\Delta_1\eta_1^2({\bf k}) (V_1W_1\Delta_1\Delta_{t1}\eta_1^2({\bf k}) 
\right. $$ 
$$\left. \hspace{2cm} +V_2W_2\Delta_2\Delta_{t2}\eta_2^2({\bf k})
-\mu Ab_0 )+\frac{\gamma}{\alpha ({\bf k})}\epsilon_{\bf k}^2W_1\Delta_{t1}
\eta_1^2({\bf k}) \right \}\frac{1}{E_{\gamma}(\bf k)}tanh
\frac{E_{\gamma}(\bf k)}{2T} 
\eqno(11)$$
\noindent and

$$\Delta_{t2}  =  \frac{1}{2N} \sum_{\bf k}\sum_{\gamma=+,-}
\left \{  W_2\Delta_{t2}\eta_2^2({\bf k})+\frac{\gamma}{\alpha(\bf k)}
V_2\Delta_2\eta_2^2({\bf k}) (V_1W_1\Delta_1\Delta_{t1}\eta_1^2({\bf k}) 
\right. $$ 
$$\left. \hspace{2cm} +V_2W_2\Delta_2\Delta_{t2}\eta_2^2({\bf k})
-\mu Ab_0 )+\frac{\gamma}{\alpha ({\bf k})}\epsilon_{\bf k}^2W_2\Delta_{t2}
\eta_2^2({\bf k}) \right \}\frac{1}{E_{\gamma}(\bf k)}tanh
\frac{E_{\gamma}(\bf k)}{2T} \eqno(12)$$

\noindent The energy eigenvalues $E_{\gamma}({\bf k})$ used above are 
$$E_{\gamma}({\bf k})=[b_m^2+|\Delta_s({\bf k})|^2+|\Delta_t({\bf k})|^2
+\epsilon_{\bf k}^2+\mu^2+2\gamma \alpha({\bf k})]^{1/2}\eqno(13)$$

\noindent where $\alpha({\bf k})=\{ (V_1W_1\Delta_1\Delta_{t1}
\eta_1^2({\bf k})+V_2W_2\Delta_2\Delta_{t2}\eta_2^2({\bf k})-\mu b_m)^2
+\epsilon_{\bf k}^2(\mu^2+|\Delta_t({\bf k})|^2)\}^{1/2}$ with $W_{1,2}$
set to zero in equations (6)-(10). 
In order to obtain the phase diagram in the temperature-density
plane, the particle density $n$ is calculated from $n=-\frac{\partial F}
{\partial \mu}$ as 

$$n  =  1+\frac{1}{2N} \sum_{\bf k}\sum_{\gamma=+,-}
\left \{  \mu+\frac{\gamma}{\alpha(\bf k)}
\left [ (-Ab_0)(V_1W_1\Delta_1\Delta_{t1}\eta_1^2({\bf k}) 
\right.  \right. $$
$$ \left. \left. \hspace{2cm} +V_2W_2\Delta_2\Delta_{t2}
\eta_2^2({\bf k})-\mu Ab_0)+\epsilon_{\bf k}^2 
\mu \right ] \right \}\frac{1}{E_{\gamma}(\bf k)}tanh
\frac{E_{\gamma}(\bf k)}{2T}\eqno(14)$$ 

\noindent Note that $\xi_{\bf k}+\xi_{\bf k+Q}=0$ when $t^{\prime}=0.$ 
The $\bf k$-sums run over half the Brillouin zone to accommodate the zone 
folding due to AF state.  
\vspace{0.4cm}

\noindent {\bf III. Results and Discussion} 
\vspace{0.4cm}

It is straightforward to check that these self-consistency equations
reduce to simpler and well known forms in the limit of pure phases
(either AF or SC). Setting $\Delta_{t}$ and $\Delta_{s}$ zero in equation
(7), we recover $E_{\gamma}({\bf k})=-\mu-\gamma\sqrt{(b_{m}^{2}+
\epsilon_{\bf k}^{2})}$ (where $\gamma,$ as before, is $\pm 1$) 
and $\alpha({\bf k})=\sqrt{(b_{m}^{2}+\epsilon_{\bf k}^{2})}.$  This leads to 
the well known self consistency equation for AF order parameter 
$1/A=-\frac{1}{2} \sum_{{\bf k},\gamma=\pm 1}\frac{\gamma tanh 
\beta E_{\gamma}({\bf k})/2} {\alpha({\bf k})}.$ Similar reduction occurs
in the equations for singlet or triplet SC order parameters in the absence
of other two.
The complete solutions of the non-linear coupled set of equations (6)-(11)
have been obtained numerically. Before discussing these solutions and
the resulting phase diagrams, we examine some of the equations critically.  

The structure of these self-consistency equations for the order parameters 
lend themselves to some interesting conclusions as noted earlier. The amplitude
$\Delta_{t1}$ has a finite value even when the pairing interaction
in the triplet channel with corresponding symmetry is zero,
provided the AF ($b_0$) and the $d_{x^2-y^2}$ SC order parameters 
($\Delta_1$) are non-zero simultaneously. In exactly similar manner 
$\Delta_{t2}$ gets {\it dynamically generated}  when both AF and $d_{xy}$ SC 
order parameters (i.e., $b_0$ and $\Delta_2$) appear simultaneously while 
the pairing interaction($W_2$) is zero in eqn.(10). Simultaneous presence of 
AF and singlet $d+id$ SC state dynamically generates the triplet $d+id$ SC 
state. This is a reflection of the fact that the presence of 
spin density wave order parameter $<c_{{\bf k},\uparrow}^{\dag}
c_{{\bf k+Q},\uparrow}>$ and singlet SC order parameter 
$<c_{{\bf k},\uparrow}^{\dag} c_{{\bf -k},\downarrow}^{\dag}>$ can lead 
to a coupling in the triplet channel $<c_{{\bf k+Q},\uparrow}^{\dag}
c_{{\bf -k},\downarrow}^{\dag}>$. Note that the symmetries of the 
order parameters for singlet and triplet SC states have to be the same. 
A glance at the terms causing the dynamical generation of triplet SC order 
parameters in eqns. (9) and (10) reveals that they contain a factor 
$\eta_i^2({\bf k})$ of which one $\eta_i({\bf k})$ term comes from spin singlet 
amplitude and the other comes from the spin triplet term. For the dynamical 
generation of triplet SC order parameters, it is necessary that both of them 
have the same symmetry or at least non-orthogonal. At half filling ($\mu=0$)
these terms responsible for the dynamical generation of triplet amplitudes
vanish and there will be no triplet SC state in the absence of $W_1$ or 
$W_2$.  
 
The self-consistency equations for the order parameters are solved numerically 
for different values of the interaction strengths $V_1$, $V_2$, $W_1$ 
and $W_2$ in the presence of nearest ($t^\prime=0$) and next nearest 
neighbour hopping ($t^\prime \ne 0$). The corresponding phase diagrams
are shown in Figs.1-5 in the doping ($x=n-1$), temperature(T) plane. 
In this calculation, all energies and temperatures are scaled  
in units of $t$. In Fig.1, where $V_1=-0.17$, $V_2=-0.08$, $W_1=W_2=0$ 
the ground state is antiferromagnetic at half filling for $t^\prime=0$.
As we move slightly away from half filling, a phase appears where 
the order parameters corresponding to AF, $d_{x^2-y^2}$-SC and the 
$\pi$-triplet SC with $d_{x^2-y^2}$ symmetry are simultaneously present. 
Note that the triplet SC phase appears even though the pairing potential  
in the triplet channel ($W_1$) is zero. This phase is generated $\it 
dynamically$ in the presence of the other two phases. 
In the region where different ordered phases coexist, the system actually 
phase separates\cite{mein}: there is a first order transition between the
SC and the AF states. In the mean-field theory there is a single 
phase boundary that separates the two phases, whereas in an actual system, 
with long range interactions, there could be multiple domains of one phase 
in another. Finally, far away from half filling we get an SC-only phase having 
$d_{x^2-y^2}$ symmetry. 

Fig.2 describes the phase diagram with a larger value of $V_2=-0.32$ 
keeping the other parameters same as in Fig.1. The phase diagram has
the same topology as in Fig.1, but an increased $V_2$ favours the $d_{xy}$
state over the $d_{x^2-y^2}$ and hence the three phase region now comprises of
AF, $d_{xy}$-SC and the $\pi$-triplet SC having $d_{xy}$ symmetry .
As before, the triplet SC phase exists (with a different symmetry compared
to Fig.1, being forced by the $d_{xy}$ symmetry of the corresponding singlet 
phase now) even without the pairing interaction $W_2$. 

The generic phase diagram of the model remains similar to either Fig.1
or Fig.2 as the ratio $\frac{V_1}{V_2}$ is changed. As noted 
earlier\cite{beck}, simultaneous appearance of an AFM phase with $d+id$
SC can be observed in a narrow regime of parameter space with strongly
suppressed AFM region. In a similar vein, we observe in Figs.3-5, a
non-zero value of both AFM and $d+id$ order parameter (along with the
dynamically generated triplet amplitude), when the interaction strength
corresponding to AFM amplitude is suppressed ($A=zV_1$). The x-T phase 
diagram for $V_1=-0.21$, $V_2=-0.32$ with $W_1=W_2=0$ and $t^\prime=0$ 
is shown in Fig.3a. At half filling the ground state is 
antiferromagnetic as usual. On increasing the filling slightly a phase
appears where AF, $d_{x^2-y^2}+id_{xy}$ SC and the triplet $d_{x^2-y^2}
+id_{xy}$
SC amplitudes are simultaneously non-zero. The triplet superconducting 
amplitude is generated $\it dynamically$ and has the same symmetry as of
the singlet SC state as expected. On increasing the doping we get a
singlet $d_{x^2-y^2}+id_{xy}$ SC phase. Note that at a higher temperature
in the phase diagram, there is a region where only the pure $d_{xy}$ phase
survives. The same phase diagram can be drawn in the temperature-chemical
potential plane (Fig.3b), a situation that obtains in some experiments where
it is difficult to dope the system while as a function of pressure there
are interesting phase transitions observed. It is quite interesting
to note the similarity between the phase diagram shown in Fig.3b here 
with Fig.1 in Lefebvre et. al.\cite{lefeb}. The symmetry of the SC phase 
abutting the AF phase in our model depends on the values of the parameters
$V_1$ and $V_2$. A dynamical generation of triplet pairing amplitude,
therefore, remains a distinct possibility in the region of coexistence 
of AF and SC phases in the organic superconductors and further experiments 
are needed to ascertain this. We did not extend our study of the model to
the high temperature normal state properties and it remains to be observed if 
strong pairing fluctuations render the single particle properties of 
that state unusual. 
  
In the phase diagram shown in Fig.4 pairing potentials for $\pi$ triplet SC
state are taken to be finite. The topology of the phase diagram remains
the same as in Fig.3. The phase boundaries shift to provide a larger triplet
region only and no separate triplet SC region arises even if the values 
of $W_1$ and $W_2$ are made comparable to those of $V_1$ and $V_2$. 

In the phase diagram of Fig.5 we introduced a small $t^\prime=0.03$ without 
changing $V_1$ and $V_2$ and keeping $W_1=W_2=0$. Changing the band structure 
with a non zero $t^\prime$ is known\cite{maitra} to favour the $d_{xy}$ 
state over the $d_{x^2-y^2}$ state. But simultaneously, the chemical potential
shifts for a particular doping on introduction of $t^\prime$. These two
effects act counter to each other in the term $\xi_{\bf k}+\xi_{\bf
k+Q}$ (which is $-2\mu$ if $t^\prime=0$). As a result, we get a small sliver of
pure $d_{x^2-y^2}$ component before the $d+id$ phase and the coexistence
region therefore has AF, $\Delta_1$ and the dynamically generated $\Delta_{t1}$.  
At high temperature we get a pure $d_{xy}$ SC phase, below which a  
$d_{x^2-y^2}+id_{xy}$ state appears.  
On increasing the next near neighbour hopping, the $d_{xy}$ phase stabilises 
and the sliver of $d_{x^2-y^2}$ state disappears (Fig.6). There is a combined
region of AF and $d+id$ state in both singlet and triplet channel. The triplet
part is, of course, dynamically generated here as $W_1=W_2=0$. For the purpose
of demonstration, we have shrunk the region of pure AF phase around half 
filling in Fig.6. We also observe that in our model the on-site Coulomb
interaction term has been set to zero. Presence of this term, at least at
the mean-field level, will not change the phase diagram 
qualitatively\cite{kyung}; although the region of stability 
of the AF phase increases with such a term.

In conclusion, we have demonstrated the possibility of simultaneous 
presence of a spin density wave and superconductivity in the $d+id$ channel in 
a model with only extended pairing interactions. As it turns out that such 
simultaneous appearance of AF and singlet SC order parameter leads to a 
spontaneous generation of a triplet SC amplitude with the same symmetry 
even when the corresponding pairing interaction in the triplet channel 
is absent. Though we do not intend to propose the present model for the
organic superconductors, the nature of phase diagrams we obtained from a
generic correlated electronic model that produces coexistence of AF and
singlet SC state bears
similarity to the ones obtained for them. Since the dynamical
generation of the triplet amplitude rests only on the coexistence of AF and
singlet SC order, we believe it would be interesting to see if further 
measurements in the organic superconductors like 
$\kappa$-(ET)$_{2}$Cu[N(CN)$_{2}$]Cl reveal the presence of unconventional
superconductivity in the triplet channel as well. 
\vspace{0.4cm}

\noindent{\bf Acknowledgement} 
\vspace{0.4cm}

AT acknowledges hospitality during the fall of 1999 from University 
of Neuchatel where this work began and TM acknowledges useful exchange
of views over email with B. Kyung. AT also acknowledges helpful discussion
with K. Behnia on the $d+id$ state.

\newpage
\center {\Large \bf Figure captions}
\vspace{0.5cm} 
\begin{itemize} 

\item[Fig. 1.] Phase diagram in doping $x=n-1$ and temperature (T) plane
for $V_1=-0.17$, $V_2=-0.08$, $W_1=W_2=0$ and $t^\prime=0$. All energies are 
measured in units of $t$.
 
\item[Fig. 2.] Phase diagram in doping and temperature plane for $V_1=-0.17$,
$V_2=-0.32$, $W_1=W_2=0$, $t^\prime=0$.

\item[Fig. 3.] a) Appearance of the dynamically generated triplet state 
for $V_1=-0.21$, $V_2=-0.32$, $W_1=W_2=0$, $t^\prime=0$. All five 
amplitudes are non-zero in the region of coexistence. In b) is shown
the same phase diagram in the T-$\mu$ plane. The dashed line represents a
first order transition, while the solid lines stand for second order 
transitions.
 
\item[Fig. 4.] No major change appears in the phase diagram with 
$W_1=-0.21$ and $W_2=-0.32$ in comparison to Fig. 3. The triplet
phase occupies slightly larger region in the phase diagram.  

\item[Fig. 5.] A non-zero small $t^\prime=0.03$ changes the phase diagram
with the sliver of $d_{x^2-y^2}$ SC state appearing in between with
no $d_{xy}$ state in the coexistence region. Both $W_1$ and $W_2$ have 
been kept zero.

\item[Fig. 6.] A larger $t^\prime=0.1$ gives rise to the region
of coexistence of all five amplitudes again. The $d_{xy}$ component
stabilises on increasing $t^{\prime}$. 

\end{itemize} 

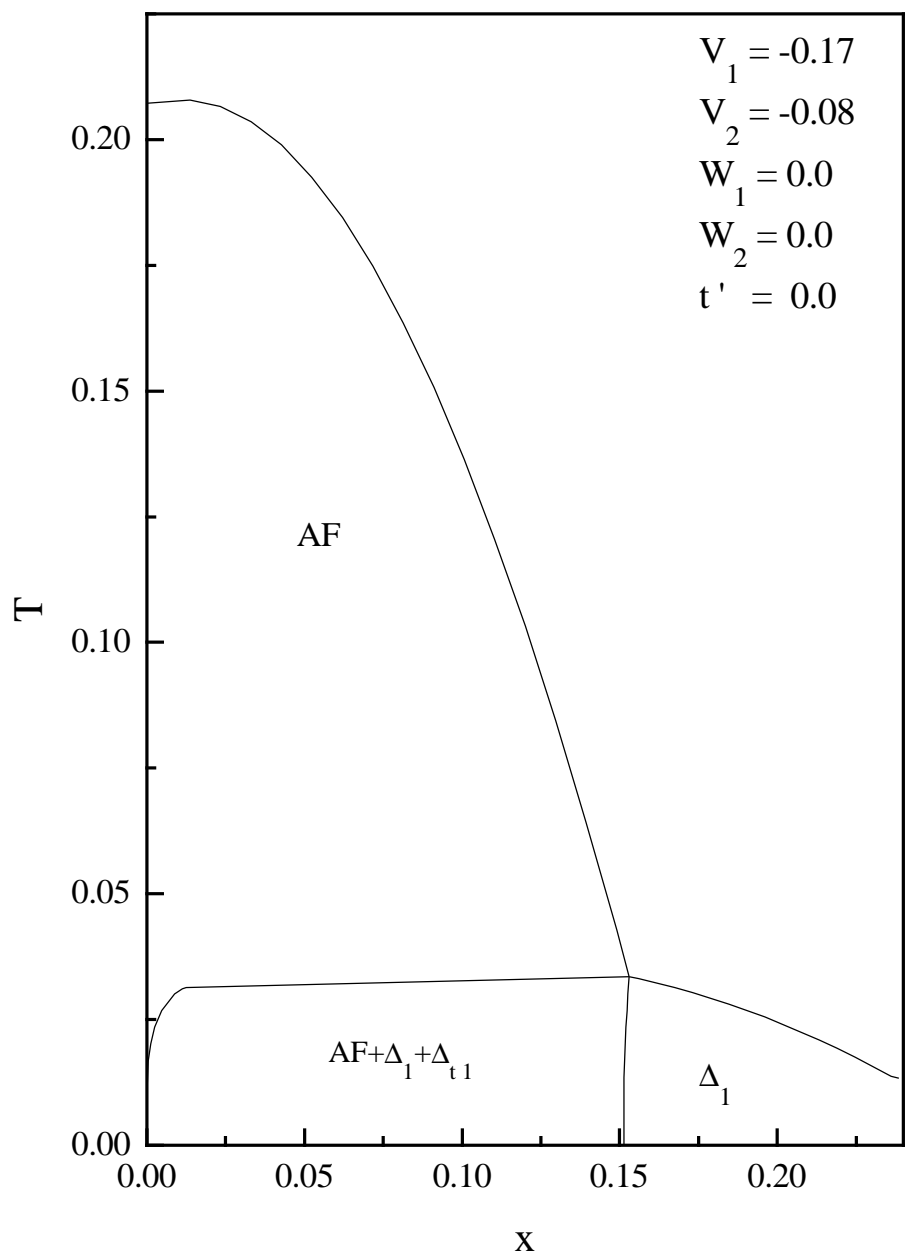
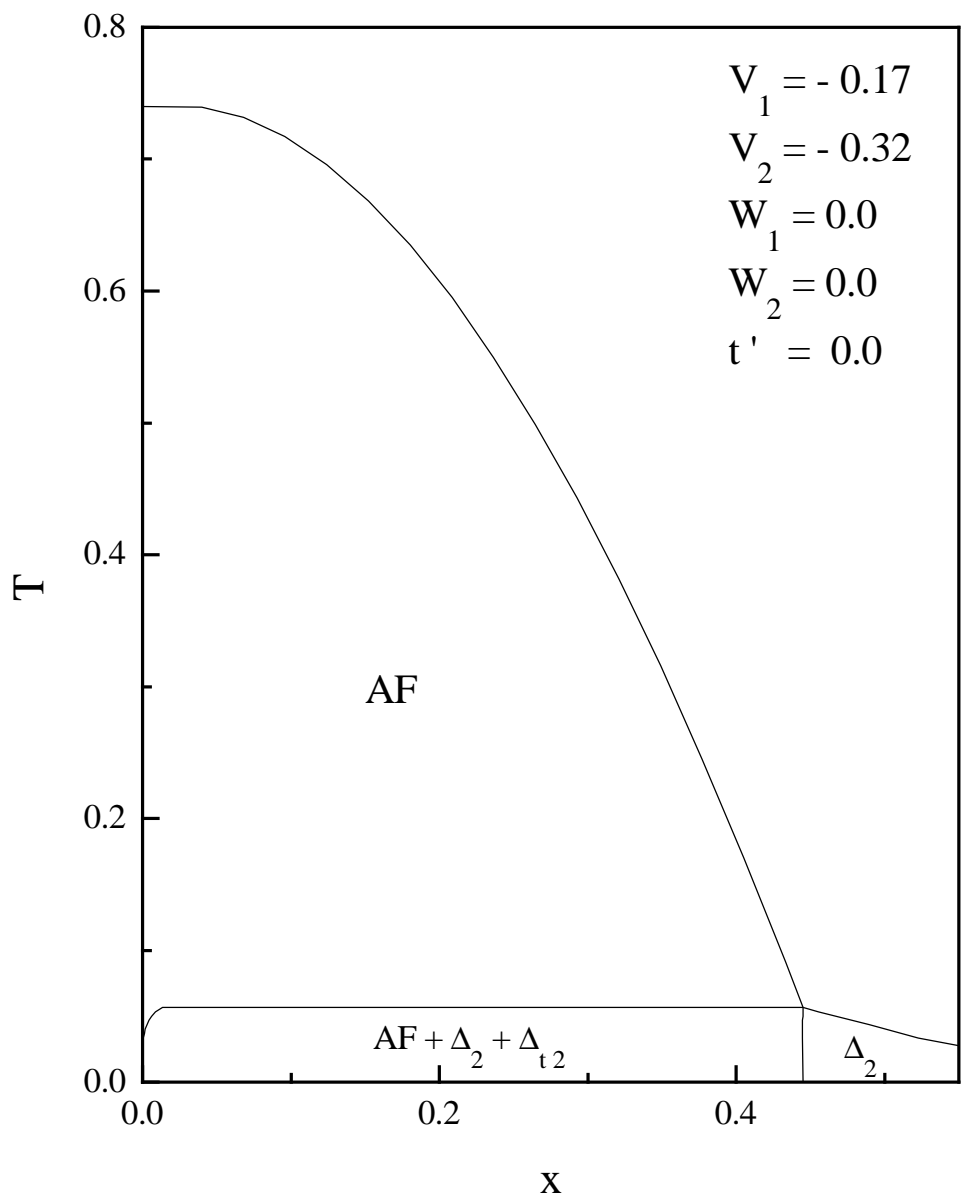
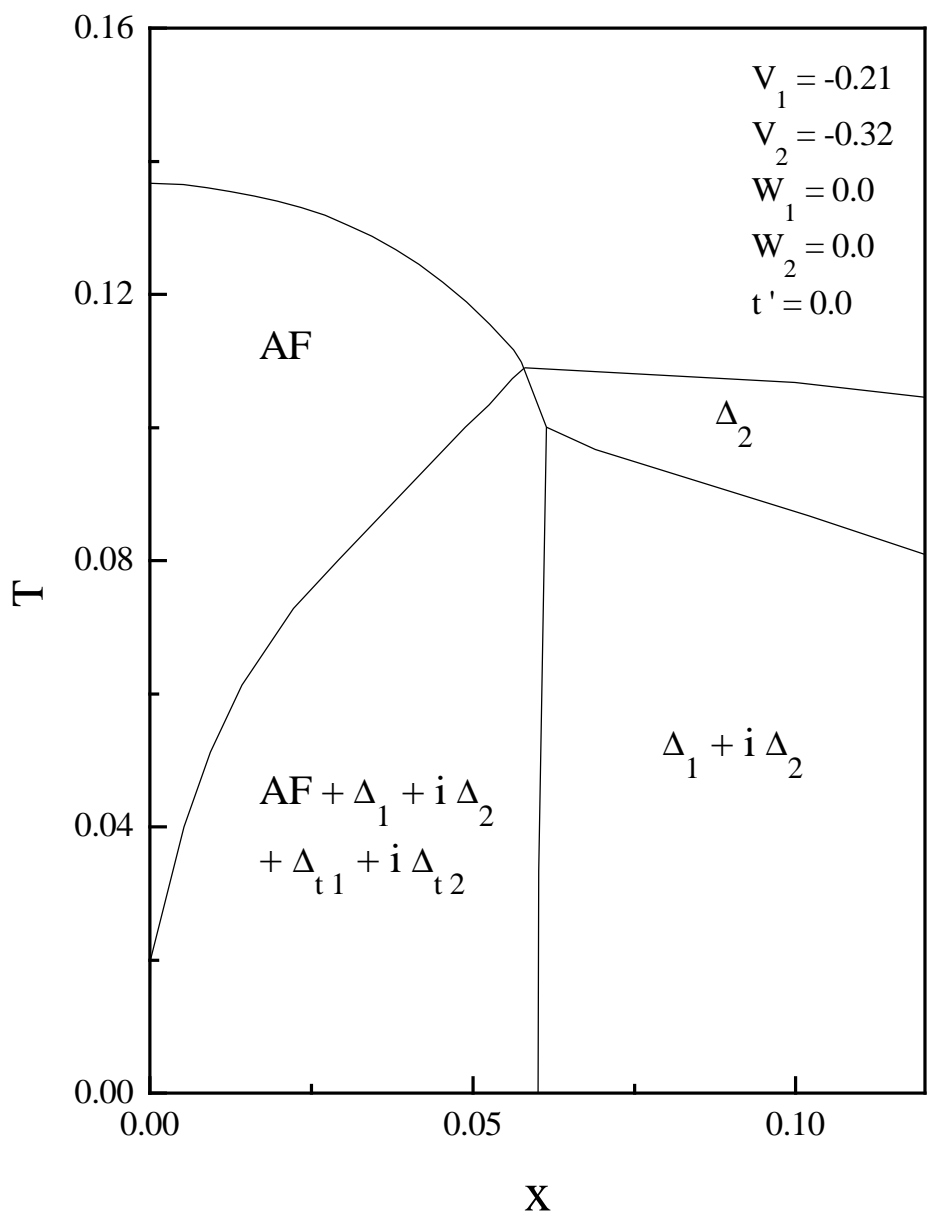
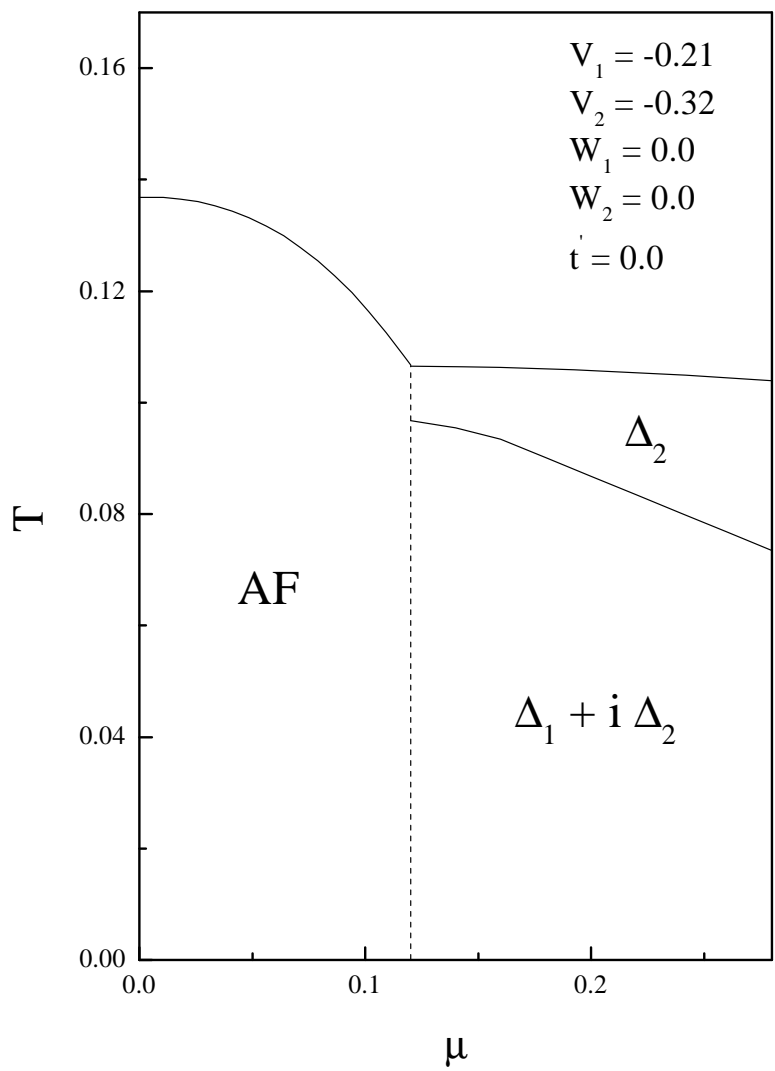
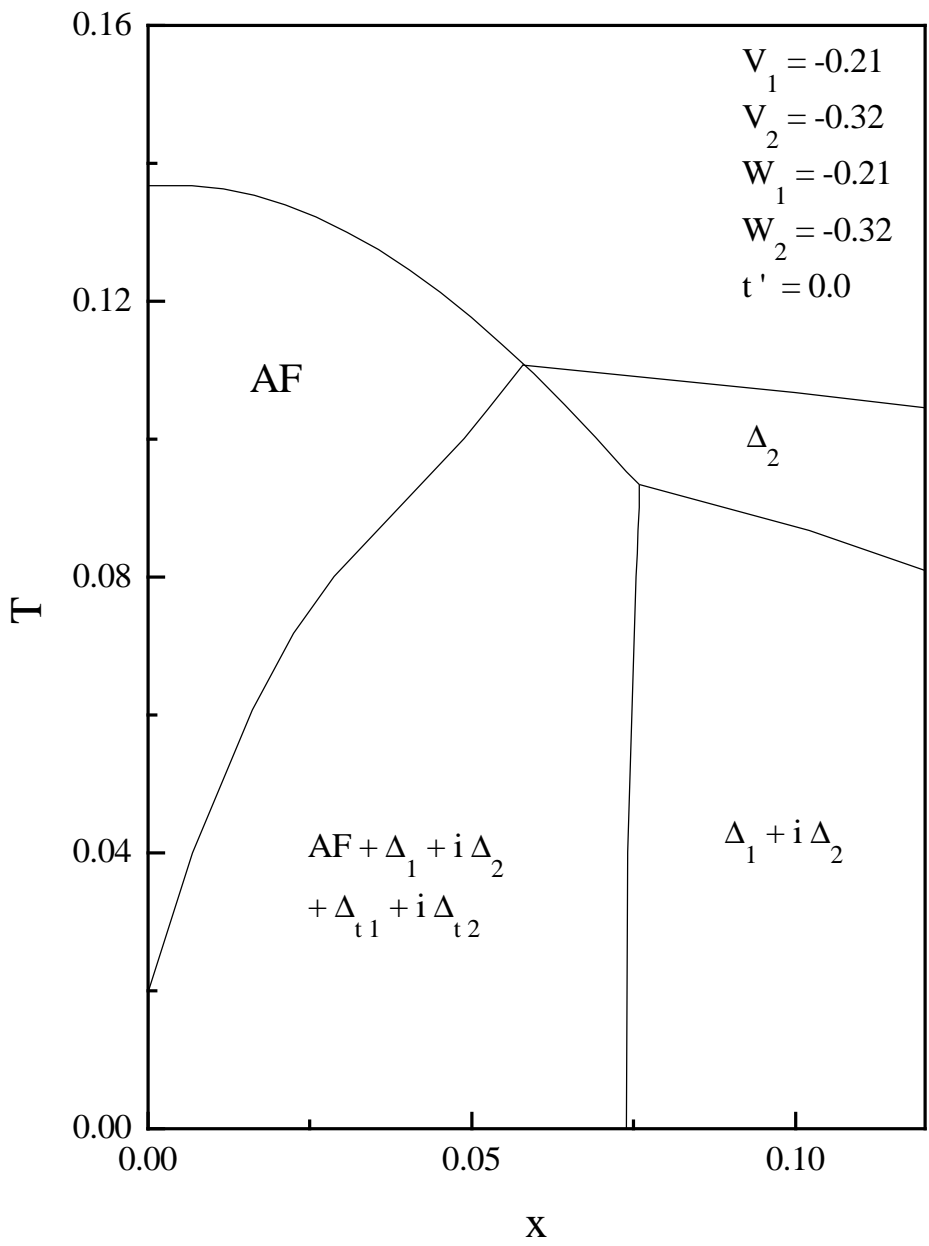
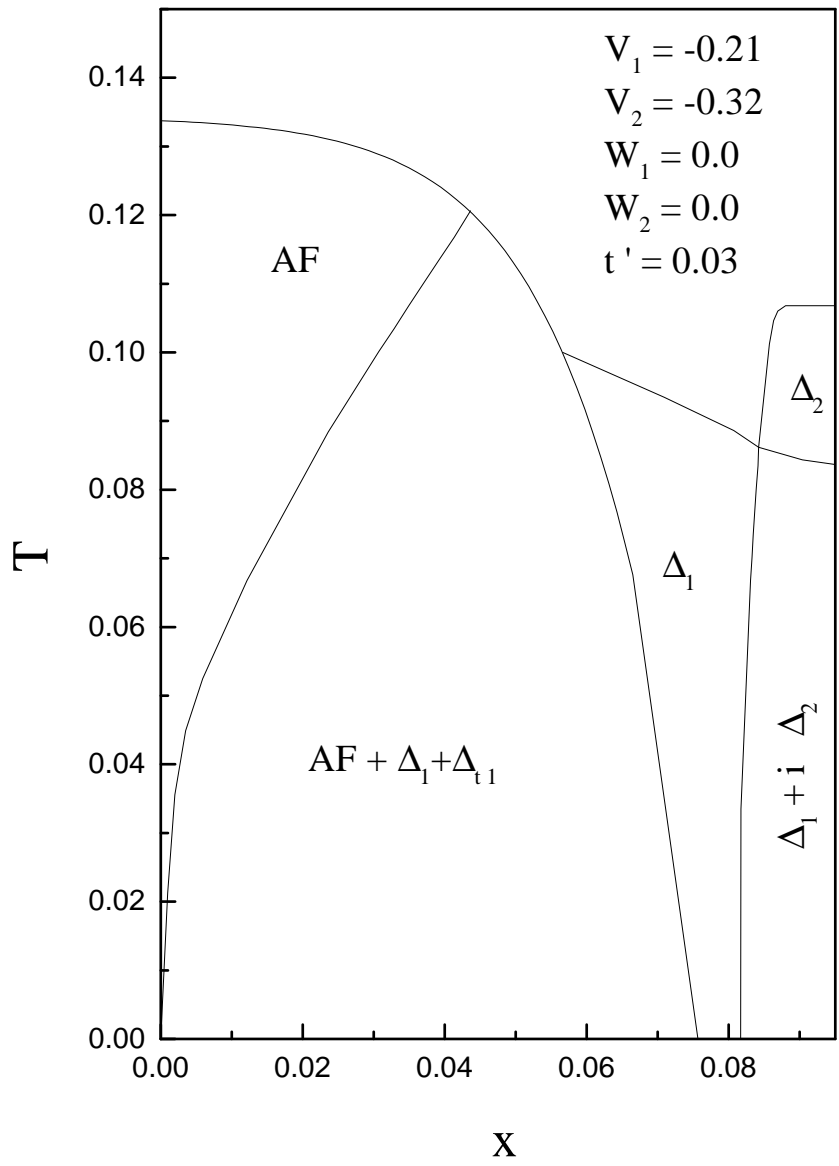
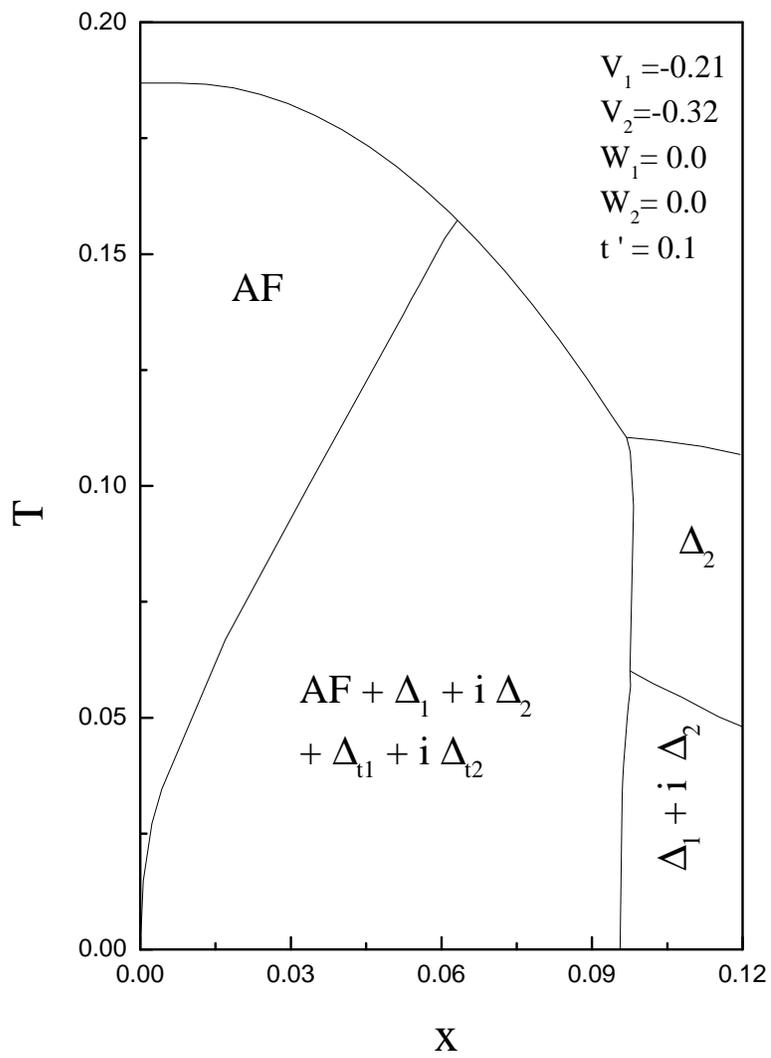
\end{document}